# Space-Confined Solid-Phase Growth of Two-Domain 1T'-ReSe$_2$ for Tunable Optoelectronics

*Yunhao Tong,[†] Fanyi Kong,[†] Lei Zhang,[†] Xinyi Hou,[†] Zhengxian Zha,[†] Zheng Hao,[⊥] Jianxun Dai,[†] Changsen Sun,[†] Jingfeng Song,[§] Huolin Huang,[†] Chenhua Ji,[‡] Lujun Pan,[§] and Dawei Li[†,]\**

[†] School of Optoelectronic Engineering and Instrumentation Science, Dalian University of Technology, Liaoning 116024, China

[⊥] Dalian University of Technology and Belarusian State University Joint Institute, Dalian University of Technology, Liaoning 116024, China

[§] Department of Materials Science and Engineering, University of Connecticut, Storrs, 06269, USA

[‡] Department of General Medicine, Dalian Municipal Central Hospital Affiliated Dalian University of Technology, Dalian 116033, China

[§] School of Physics, Dalian University of Technology, Dalian 116024, China

\* Address correspondence to: dwli@dlut.edu.cn

**KEYWORDS**: *1T'-ReSe$_2$, low-temperature growth, two-domain structure, carrier transport modulation, phototransistors*

**ABSTRACT**: Two-dimensional layered ReX$_2$ (X = Se, S) has attracted researcher's great interest due to its unusual in-plane anisotropic optical and electrical properties and great potential in polarization-sensitive optoelectronic devices, while the clean, energy-saving, and ecological synthesis of highly-crystalline ReSe$_2$ with controlled domains remains challenging yet promising. Here, we develop a novel space-confined solid-phase approach for the growth of high-quality two-domain 1T'-ReSe$_2$ with tunable optoelectronic properties by using pure Re powder film as Re precursor. The results show that ReSe$_2$ can be grown at a temperature as low as 550 °C in a small-tube-assisted space-confined reactor, with its size and shape well-tailored via temperature control. A solid-phase two-domain ReSe$_2$ growth mechanism is proposed, as evidenced by combining *in-situ* optical monitoring, *ex-situ* electron microscope and elemental mapping, and polarized optical imaging. Moreover, we have fabricated two-domain ReSe$_2$ transistors, which exhibit switchable transport behavior between *n*-type and ambipolar character via grain boundary orientation control. This modulation phenomenon is attributed to the different doping levels between the grain boundary and the single domain. Furthermore, the as-fabricated two-domain ReSe$_2$ photodetectors exhibit a highly gate-tunable current on-off ratio (with a maximum value of ~8.2×10$^3$), a polarization-sensitive photo-response, and a high-speed response time (~300 μs), exceeding most of the previously reported ReX$_2$ photodetectors. Our work thus provides a new, low-consumption, energy-saving growth strategy toward high-quality, domain-controlled ReX$_2$ for highly tunable and high-performance optoelectronics.

**INTRODUCTION**

Two-dimensional (2D) layered ReX$_2$ materials, such as ReS$_2$ and ReSe$_2$, exhibit unusual in-plane anisotropic optical and electrical properties due to their unique structure,[1-4] which endows them with great potential applications in polarization-sensitive photodetectors,[5-7] nanoscale polarization controllers,[8] logic field-effect transistors (FETs),[4,9] integrated digital invertors,[2] and optical logic circuits.[10] Both ReS$_2$ and ReSe$_2$ have the same distorted octahedral (1T') crystal structure with triclinic symmetry, where the diamond-shaped Re 4-clusters are formed along the *b*-axis (Figure 1a). However, they exhibit different electrical properties. For example, ReS$_2$ is a direct bandgap semiconductor, with the bandgap ($E_g$) varying from 1.43 eV in monolayer to 1.35 eV in bulk,[11] in contrast to the direct in bulk (1.38 eV) to indirect in bilayer (1.73 eV) and again to direct in monolayer (2.05 eV) in ReSe$_2$.[12] In addition, ReS$_2$ behaves as an *n*-type semiconductor,[4,13] while ReSe$_2$ has been shown to possess an ambipolar electronic character.[6,14] Although great progress has been made in ReS$_2$ from growth to device applications,[15] ReSe$_2$ is dramatically lagging behind in these aspects, especially in the controllable fabrication of ReSe$_2$ with tunable optoelectronic performances.

To date, several methods have been developed to produce ReSe$_2$ films, including mechanical exfoliation, physical vapor deposition (PVD), and chemical vapor deposition (CVD).[15] Whereas the high-quality ReSe$_2$ can be produced by mechanical exfoliation and PVD methods, the mechanical exfoliation is limited by low productivity, uncontrolled domain size and layer thickness,[10,16] and PVD requires ultrahigh growth temperature (900-1000 °C) and high vacuum environment,[17] thus hindering these methods for the practical applications. In contrast, CVD has become a popular approach for cost-effective and scalable growth of ReX$_2$ crystal layers.[13,18-20] In most of the previous reports,[21-24] the ReSe$_2$ grown by CVD is usually a polycrystalline material containing randomly distributed grain boundaries (GBs). Although the electrical properties of 2D materials could be tailored via GBs,[25-26] the CVD growth of ReSe$_2$ with controlled GBs remains challenging yet promising. Additionally, ReO$_3$ powder is usually used as Re precursor and dangerous H$_2$ carrier gas is required for the CVD growth of ReSe$_2$, which may introduce oxygen element impurity and affect the quality of as-grown products. Recently, pure Re powder has been reported to fabricate high-quality ReS$_2$, which is cleaner and cheaper than using ReO$_3$, but still requires a relatively high temperature (~750 °C).[27] We also note that the ReX$_2$ growth dynamics based on as-reported CVD method belong to gas-phase reaction, namely, Re element participates the reaction in vapor state, making it disposable for the Re precursor. A possible approach to increase the Re precursor utilization is developing a solid-phase CVD reaction system,[28] which may enable the clean, low energy-cost, and ecological growth of highly-crystalline ReSe$_2$ with controlled domains, but has never been explored to date.

In this work, we report a facial space-confined solid-phase strategy to the low-temperature growth of highly-crystalline 1T'-ReSe$_2$, which has been demonstrated to have controlled GBs and tunable optoelectronic properties. The results show that ReSe$_2$ samples can be grown at a temperature as low as 550 °C by using pure Re powder as Re precursor in a small-tube-assisted space-confined reactor. In addition, ReSe$_2$ flakes become larger in size as the temperature increases and the shape of ReSe$_2$ is mainly triangle (strip) at low (high) temperature, suggesting that the size and shape can be well tailored via temperature control. Moreover, a solid-phase two-domain ReSe$_2$ growth mechanism is proposed, as evidenced by combining *in-situ* optical monitoring, *ex-situ* scanning electron microscope (SEM) and energy dispersive X-ray spectral (EDX) mapping, and polarized optical imaging. Furthermore, we have fabricated two-terminal FETs using as-grown two-domain ReSe$_2$ with one GB, and find that the control of GB orientation relative to the channel can significantly modulate the transport behavior from *n*-type (parallel case) to ambipolar (perpendicular case) electrical character, which is attributed to the different doping levels between the GB and the single domain. Moreover, two-domain ReSe$_2$ device as a photodetector exhibits a highly gate-tunable current on-off ratio with a maximum value of ~8.2×10$^3$ and a high-speed rising/falling response time of ~0.3/0.3 ms, exceeding most of the previously reported ReX$_2$ photodetectors. Overall, this study provides a clean, low-cost, energy-saving strategy to fabricate high-quality two-domain anisotropic semiconductors with controlled GBs, paving the way for the development of 2D-based highly-tunable and high-performance phototransistors.

**RESULTS AND DISCUSSION**

**Low-temperature growth of 1T'-ReSe$_2$ via space-confined strategy.** 1T'-ReSe$_2$ is a layered low-symmetry semiconductor with its *b*-axis parallel to Re chain direction (Figure 1a). Herein, we design a space-confined reactor (SCR) for high-quality ReSe$_2$ growth by using pure Re and Se powder as precursors (Figure 1b), where the top substrate/Re powder film/bottom substrate (T-sub./Re/B-sub.) sandwiched structure is placed at the center of CVD system with one heating zone, while the Se powder is put at the edge of the heating zone. Although Re powder remains in solid state during reaction process owing to its ultrahigh thermal stability (Supporting Information), we find that it could be used as efficient precursor for ReSe$_2$ growth at low temperature via our space-confined strategy. As shown in Figure 1c, after reaction in SCR at 600 °C, triangle-shaped ReSe$_2$ flakes are formed on both top and bottom SiO$_2$/Si substrate surfaces.

Interestingly, when the top substrate is removed from T-sub./Re/B-sub. sandwiched structure, no ReSe$_2$ is grown using the same CVD process (Supporting Information), which reveals the crucial role of top substrate in building a space-confined environment. The similar phenomenon has been previously reported for the growth of ultrathin MoS$_2$ single crystals via space-confined strategy.[28] It is considered that, in our SCR system, Se gas enters the porous structure of Re powder film between two substrates, which provides a long time for Se and Re reaction at the Re powder/substrate interface (Figure 1b inset). If the top substrate is removed, Se gas will escape from the surface of Re powder film in a short time, resulting in little contact between the Se gas and the Re powder, thereby explaining why ReSe$_2$ could not be formed.

Figure 1d and 1e show the atomic force microscope (AFM) images of two typical as-grown triangle-shaped ReSe$_2$ flakes, which have a uniform layer thickness of ~6 nm and ~13 nm, respectively. Polarized optical imaging characterization (Figure 1d and 1e insets) suggests that these triangle-shaped samples are composed of two domains.[29] In addition, a ~60° angle in each triangle-shaped flake is defective (red arrows in Figure 1d and 1e). These observations indicate a unique growth mechanism for ReSe$_2$ via space-confined strategy, as will be discussed later. Figure 1f shows the Raman spectrum for as-grown ReSe$_2$ flakes, where in-plane $E_g$-like mode (at 124 cm$^{-1}$) and out-of-plane $A_g$-like mode (at 160 and 173 cm$^{-1}$) are detected, consistent with the previous reports.[30]

Next, we investigate the effect of temperature on ReSe$_2$ growth via space-confined strategy. Figure in Supporting Information shows typical optical images of ReSe$_2$ samples grown at 550, 600, 650, 700, 750 and 800 °C. We find that, using SCR reactor (Figure 1b), the lowest temperature required for ReSe$_2$ growth is around 600 °C. As the temperature increases, ReSe$_2$ flakes become larger and thicker. In addition, ReSe$_2$ flakes with different shapes are produced at each temperature, mainly including triangle, trapezoid, parallelogram, and strip. Figure 1g shows the relative shape proportion for ReSe$_2$ grown at different temperatures. A tendency is that the shape of as-grown ReSe$_2$ is mainly triangle (strip) at low (high) temperature, which can be attributed to the difference of edge formation energy for different shaped ReSe$_2$.[31] When the feature size of each shape is defined (Figure 1g insets), we draw out the temperature dependence of average feature size ($L_a$) for triangle-/trapezoid-/parallelogram-/strip-shaped ReSe$_2$ (Figure 1h), all of which show a positive linear correlation between $L_a$ and temperature. Hence, both the size and the shape of SCR ReSe$_2$ could be well tailored via temperature control.

To study the effect of micro-environment around T-sub./Re/B-sub. sandwiched structure on ReSe$_2$ growth, we design a small tube-assisted space-confined reactor (ST-SCR) (Figure 2a). Considering the lower surface energy of mica compared to SiO$_2$,[32] mica and SiO$_2$/Si slices are used as substrates for comparison. It is found that the growth of ReSe$_2$ using SiO$_2$/Si substrate in ST-SCR (Figure 2b, top) is more productive than that in SCR (Figure 1c). The similar results are obtained on SiO$_2$/Si (Figure 2b, top) and mica (Figure 2b, bottom) substrates, showing a universality of growth effect on different substrates. More importantly, the growth of ReSe$_2$ in ST-SCR can be realized at a very low temperature (≤ 550 °C), which is impossible in SCR. The previous studies have shown that the flow field and vapor concentration play a critical role in the growth of ReS$_2$.[33] In our case, the difference between ST-SCR and SCR is the change in flow field, which leads to local alternation of Se gas concentration, thereby influencing the growth results.

To reveal the relationship between the temperature and the ST-SCR ReSe$_2$ growth, we have analyzed the shape ratio (Figure 2c) and size distribution (Figure 2d) for the samples grown on SiO$_2$/Si and mica substrates at temperature ranging from 550 to 650 °C. As shown in Figure 2c, triangle-shaped ReSe$_2$ flakes occupy a highest proportion among four regular shapes at all temperatures. As the temperature increases from 600 to 650 °C, the triangle (strip) ratio is decreased (increased), consistent with the case in SCR. However, for the ST-SCR, no obvious correlation between ReSe$_2$ average feature size and temperature is observed (Figure 2d). In addition, the average feature size for four kinds of shaped ST-SCR ReSe$_2$ flakes is two times larger than that grown in SCR, indicating much more efficient ReSe$_2$ growth via small-tube-assisted space-confined strategy.

**Growth mechanism of ReSe$_2$ via space-confined strategy.** To investigate the growth mechanism of ReSe$_2$ flakes via space-confined strategy, an *in-situ* monitoring experiment is designed (Figure 3a). In detail, we prepare a transparent mica substrate/Re powder film/SiO$_2$ substrate sandwiched structure for ReSe$_2$ growth. After growth, the sandwiched structure is taken out for *in-situ* optical observation. Before separating the top (mica) substrate, some ReSe$_2$ flakes on mica substrate together with Re powder contacted are observed from the back side (Stage 1). Next, the top (mica) substrate is carefully separated from Re powder/SiO$_2$ substrate (Stage 2), followed by observing these flakes from the front side (Stage 3). Figure 3b and 3c show the optical images of a typical triangle-shaped ReSe$_2$ flake taken from the back side and the front side, respectively. It is clear that a cluster of Re powder resides at one vertex of ReSe$_2$ flake. Because of the tight contact between Re powder and ReSe$_2$ vertex, we think this vertex is a nucleation point. As schematized in Figure 3d, ReSe$_2$ nucleates and grows from the contact point between Re powder and substrate. This solid-phase one-vertex-nucleation behavior is different from the gas-phase center-nucleation mechanism.[34] Similar phenomenon is observed in ReSe$_2$ with different regular shapes (Supporting Information), indicating the universality of one-vertex-nucleation feature. In this way, we deduce that the angle defections in ReSe$_2$ shown in Figure 1d-1e are the nucleation points where the growth starts. The disappearance of Re powder at vertex of these samples is probably caused during the substrate separation.

To confirm the proposed growth model, we have performed SEM and elemental mapping study (Figure 3e-3h). Figure 3f shows an SEM image of a triangle-shaped ReSe$_2$ flake with Re powder at one vertex, where the powder sticks firmly to the vertex (Figure 3f inset). It is suggested that the powder is not occasionally dropped at the triangle-shaped ReSe$_2$ vertex, but continuously in contact with it during the growth process. Figure 3e compares the EDX spectra taken at ReSe$_2$ triangle center and powder position. It is extracted from Figure 3e that the atomic ratio of Re and Se at ReSe$_2$ center is about 1/2, consistent with the stoichiometric ratio of ReSe$_2$. At the powder position, the Re/Se ratio is about 3.2/2, confirming that Se gas has participated in the reaction with solid-phase Re powder at the Re/substrate interface. EDX elemental mapping shows that the distribution of Re and Se at the powder position is different (Figure 3g and 3h). The Se element mainly distributes at the edge of powder whereas the Re element is at the center, indicating that the reaction seems to occur from the edge of Re powder. Based on the above discussion, we conclude that Re powder with some Se element absorbed at the edge contacts tightly with ReSe$_2$ vertex, which acts as the nucleation point during the growth process.

To further understand the growth mechanism via space-confined strategy, we explore the ReSe$_2$ growth dynamics by analyzing representative triangle-shaped samples at different growth stages (Figure 4a). Considering that the *b*-axis of ReX$_2$ is parallel to the long edge (or strip) direction,[9, 35] the *b*-axes as well as the nucleation points for as-grown triangle-shaped ReSe$_2$ in Figure 4a are marked. Based on the experimental observation, we propose a four-step growth model to explain the growth dynamics of triangle-shaped ReSe$_2$ (Figure 4b, top schematic). Firstly (nucleation step), the ReSe$_2$ crystal forms at the nucleation point and splits into two short branches in two directions (or *b*-axis orientations) with the angle of 60°. Secondly (expansion step), two branches continue extending along the *b*-axis direction meanwhile broadening along the direction perpendicular to *b*-axis. Thirdly (merging step), the space between two branches begins merging and a GB is formed. Finally (formation step), when the space is all merged, a triangle-shaped ReSe$_2$ flake with two domains is obtained. This model can also be used to explain the growth dynamics of ReSe$_2$ of other shapes. For example, a parallelogram-shaped ReSe$_2$ flake should correspond to a four-domain structure (Figure 4b, bottom schematic).

To prove the growth dynamics model proposed above, we analyze the correlation between the nucleation point (or Re powder) and the domain structure in ReSe$_2$ via polarized optical imaging (Supporting Information). The previous studies have shown that angle-resolved polarized optical imaging can be used to determine the crystal orientation of anisotropic 2D materials.[36] Figure 4c and 4d show the polarized optical images of a triangle-shaped ReSe$_2$ sample with Re powder at one-vertex taken in parallel and vertical configurations, respectively, both of which confirm that two domains (①and ②) with a GB exist in triangle-shaped ReSe$_2$. Figure 4e and 4f shows the polar plots of parallel polarized light intensity for two domains as a function of sample rotation angle. It can be seen that the minimum value direction (double arrow) for domain ① (②) is parallel to long edge L$_1$ (L$_2$), indicating that the *b*-axis of each domain is along the edge direction (Figure 4g),[37] as further evidenced by angle-resolved polarized Raman measurement (Supporting Information).[38] Based on this rule, we have measured more triangle-shaped and parallelogram-shaped ReSe$_2$ flakes via polarized optical imaging method (Supporting Information), and the same result is obtained: two (four) domains with one (three) GBs exist in triangle (parallelogram) ReSe$_2$ and the *b*-axis of each domain is parallel to corresponding edge direction, thus confirming the proposed growth dynamics model (Figure 4b).

**Tunable electrical and optoelectronic properties in two-domain ReSe$_2$.** In this section, we study the influence of GB in the electrical and optoelectronic properties of two-domain ReSe$_2$ by fabricating two-terminal FET devices. Based on the relative direction between the conducting channel and the GB, three types of FETs are fabricated: (1) ReSe$_2$ device with GB parallel to the channel (D-1, Figure 5a), (2) ReSe$_2$ device with GB italic to the channel (D-2, Figure 5b), and (3) ReSe$_2$ device with GB perpendicular to the channel (D-3, Figure 5c). All GBs are identified using polarized optical imaging (Figure 5a-5c). The difference of these devices is that the GBs in D-1 and D-2 are connected with the electrodes, while the GB in D-3 has no intersection with the electrodes.

Next, the transfer ($I_{DS}$-$V_{BG}$) measurements are carried out (Figure 5d-5f), which reveal that all these devices can be effectively tuned by the back-gate voltage. For direct comparison, the channel conductance ($\sigma$) is calculated using the equation $\sigma = \frac{l}{w} \cdot \frac{I_{DS}}{V_{DS}}$, where $l$ and $w$ are the length and width of the channel, $I_{DS}$ and $V_{DS}$ are the source-drain current and source-drain voltage. It can be seen that both D-1 (Figure 5d) and D-2 (Figure 5e) exhibit an *n*-type behavior, a high on-state current, and a relatively high current switching ratio (~50 to ~400). In contrast, D-3 (Figure 5f) shows a bipolar behavior, a weak on-state current, and a low current on-off ratio (~7). We also find that the transport behavior of D-3 is similar to that of one-domain ReSe$_2$ without GB (Figure 5g), revealing that the GB in two-domain device without connecting the electrodes has little influence on the doping behavior of ReSe$_2$.

The GB-induced transport modulation can be attributed to the different doping levels between the GB and the domain region (Figure 5d-5f insets),[26] where the GB provides more conductive channel (red arrows) due to heavily electron doping than that of domain area (black arrows). In detail, the current in D-1 and D-2 mainly passes through the GB rather than two domains, thus resulting in GB modulation effect. The length of GB in D-2 (italic case) is longer than in D-1 (parallel case), which is the main cause for a relatively weaker on-state current and a lower current on-off ratio compared to D-1. In case of D-3 (perpendicular case), no highly electrical conduction channel is formed, thus the GB has no obvious influence on the doping behavior in ReSe$_2$. To confirm the *n*-doping effect at GB, we have performed Kelvin probe force microscopy (KPFM) measurement (Supporting Information). The results show that the work-function at GB ($\Phi_{GB}$) is decreased compared with that at single domain ($\Phi_{domain}$), and the Fermi-level shifts toward the conduction band, thus indicating of the *n*-type doping at GB. To further support our proposed model, we design a new ReSe$_2$ device with three pairs of narrow electrodes (Figure 5h), which enables accurate comparison of current contribution (or doping level) from the GB and the domain area. As shown in Figure 5h (D-4), both left and right channel are single domains (labeled as L and R), whereas the middle channel is a GB connecting the source and drain electrodes (labeled as B). Output ($I_{DS}$-$V_{DS}$) measurements show that $I_{DS}$ in channel B is over one order of magnitude higher than that in channels L and R (Figure 5i), strongly confirming that the GB is more conductive than single domain. Overall, the control of GB can significantly modulate the electrical behavior of two-domain ReSe$_2$ devices.

In Figure 6, we have studied the optoelectrical performance of two-domain ReSe$_2$ FETs with GB parallel (D-1, Figure 6a) and perpendicular (D-3, Figure 6e) to the channel by using a focused 532 nm laser as the light source. The $I_{DS}$-$V_{DS}$ curves (at $V_{BG}$ = 0 V) for D-1 and D-3 with/without the laser illustration are shown in Figure 6b and 6f, respectively. Both devices show a high $I_{on}/I_{off}$ ratio of over three orders of magnitude, demonstrating a strong photosensitivity. Figure 6c and 6g show the transfer curves ($I_{DS}$-$V_{BG}$) for D-1 and D-3 with/without the laser illustration, demonstrating a gate tunability. To quantitively evaluate the influence of gate in photo-response, $I_{on}/I_{off}$ ratio as a function of $V_{BG}$ are calculated (Figure 6d and 6h). Over the entire gating range (-40 ≤ $V_{BG}$ ≤ 40 V), a high $I_{on}/I_{off}$ ratio is obtained in both two-domain devices, which reaches ~8.2×10$^3$ (at $V_{BG}$ = -40 V) for D-1 and ~3.2×10$^3$ (at $V_{BG}$ = -25 V) for D-3, superior to or comparable with that of one-domain ReSe$_2$ device without GB (~5.0×10$^3$ at $V_{BG}$ = -20 V) (Supporting Information).

Considering the unique in-plane crystal anisotropy in two-domain ReSe$_2$, we perform polarization-dependent photo-response study. Figure 6i shows the polar plot of normalized photocurrent ($I_{ph}$) (at $V_{DS}$ = 1 V and $V_{BG}$ = 0 V) for D-1 (parallel case) as a function of incident laser polarization direction ($\theta$), where the $I_{ph}$ reaches its maximum value when the incident light polarization is close to the GB direction (white dashed line, Figure 6a) but not along the *b*-axis of each domain (red and blue double arrows, Figure 6a). This polarization-dependent photo-response can be attributed to the combined contributions from the GB and two domain areas, because both of them exhibit strong photo-responses as discussed above. In addition, we have measured the polarization-dependent photo-response in D-3 (perpendicular case) in comparation with the corresponding one-domain channels (Figure 6j-6l), all of which show strong polarization-sensitive photoresponsivity. Unlike in parallel case, the maximum $I_{ph}$ value of two-domain device in perpendicular case is not along the GB direction (white dashed line, Figure 6e) but parallel to the *b*-axis of one domain (Domain B), reflecting a tunable but complex optoelectrical behavior. The exact modulation mechanism in this case still requires further study. The polarization-dependent photo-response at different $V_{BG}$ is also measured (Supporting Information), which shows a gate-insensitive polarization-dependent photoresponsivity in both two-domain and one-domain devices.

Moreover, we have measured the time-resolved photo-response with the laser periodically switched on and off by a chopper (Figure 6m). The device (D-3) exhibits a stable response to the laser illustration. The photo-response speed of the device is usually characterized by the rising (falling) time in photocurrent upon turning on (off) the incident light. Here, our device shows a high-speed response in both rising and falling edge (Figure 6n-6o), which are ~0.3 ms and ~0.3 ms, respectively. Finally, we compare the optoelectronic performance of our two-domain ReSe$_2$ device with other reported ReX$_2$-based photodetectors.[18, 21, 32, 39-43] It can be seen from Figure 6p that the ~8.0×10$^3$ $I_{on}$/$I_{off}$ ratio and the ~0.3 ms response time achieved in our work exceed most congener reports, confirming the realization of tunable and high-performance polarization-sensitive photodetectors based on two-domain ReSe$_2$.

## CONCLUSIONS

In summary, we have realized low-temperature (≤ 550 °C) growth of highly-crystalline ReSe$_2$ with pure Re powder via a space-confined solid-phase strategy. Using this unique method, the shape and the grain size of as-grown ReSe$_2$ could be well tailored via temperature control. We have systemically investigated the relationship between as-grown ReSe$_2$ and solid-phase Re powder as well as the ReSe$_2$ growth dynamics by combing *in-situ* optical monitoring, *ex-situ* SEM/EDX mapping and polarized optical imaging, based on which a solid-phase two-/multi-domain ReSe$_2$ growth mechanism is confirmed. In addition, we have studied the influence of GB in the electrical and optoelectronic properties of two-domain ReSe$_2$. It is found that the control of GB orientation in two-domain devices can significantly modulate the electrical transport behavior of ReSe$_2$, which is attributed to the different doping levels between the GB and the domain region. Moreover, as-fabricated two-domain ReSe$_2$ phototransistor exhibits a $V_{BG}$-tunable $I_{on}$/$I_{off}$ ratio (with a maximum value of ~ 8.2×10$^3$), an ultrafast response time (~0.3 ms), and a polarization-sensitive photo-response. Overall, our study provides a low-cost, energy-saving strategy to grow highly-crystalline 2D anisotropic semiconductors with controlled GBs, which show great potential applications for tunable and high-performance nanoelectronics and optoelectronics.

**METHODS**

**Space-confined growth of ReSe$_2$.** ReSe$_2$ thin flakes were prepared by a space-confined solid-phase growth strategy (Figure 1b). First, a substrate/Re powder film (~0.5 mm)/substrate sandwiched structure was designed and fabricated. Second, this sandwich structure was placed to the center of CVD system with one heating zone, and Se powder (~500 mg) was put at the edge of the heating zone. Next, argon (Ar) gas with a flow rate of 50 sccm was introduced, followed by heating the CVD furnace to targeted temperature (500 to 800 °C) and maintained for 20 min. After growth, the furnace was cooled down to the room temperature, and then the as-grown samples were taken out. It should be noted that the Re powder film remained could be reused as precursor for several times. To investigate the effect of micro-environment on ReSe$_2$ growth, we have altered the CVD system by adding a small quartz tube (Figure 2a), with the mouth of small tube against Ar flow direction. In this modified reactor, Se powder was placed at one end of small tube (low temperature region), while Sub./Re/Sub. sandwiched structure was placed closed to the other end (high temperature region).

**Morphological and optical characterizations.** The thickness and morphology of as-grown ReSe$_2$ were measured by AFM (Asylum Research MFP 3D, Oxford instrument) with a non-conductive tip (HQ:NSC14/Al BS) in AC mode. The structure and component of ReSe$_2$ products were detected by SEM (SU5000, Hitachi) with energy dispersive X-ray spectral detection capability. The size and shape of ReSe$_2$ samples were identified by optical microscope (Eclipse LV150N, Nikon). The composition of as-grown ReSe$_2$ samples was also identified by spectral measurement in a micro-Raman system (Renishaw InVia plus, Renishaw) using a 532 nm laser as excitation source. To identify the crystal orientation of as-grown ReSe$_2$, we performed angle-resolved polarized optical imaging and polarized Raman measurements.

**ReSe$_2$ device fabrication and characterization.** ReSe$_2$ transistor devices were fabricated by transferring as-grown samples onto SiO$_2$(300 nm)/Si substrate with pre-patterned Ti(5 nm)/Au(25 nm) electrodes,[4] where the Ti/Au electrodes were prepared by standard photolithography technique. To improve the ohmic contact, all devices were treated by thermal annealing at 120 °C for 30 min. The electrical and optoelectrical measurements were conducted at room temperature by semiconductor analyzer (B1500A, Keysight). For optical response measurement, a 532 nm laser with a power density of 12 μW/μm$^2$ was used as the light source, and a half-wave plate was placed in the optical path to control the laser polarization direction.

**ACKNOWLEDGEMENTS**

This work was supported by the National Natural Science Foundation of China (Grant Nos. 12274051, 51972039, 52272288, 61971090 and 62101093), the "Chunhui Project" Cooperative Research Project of Ministry of Education (Grant No. HZKY20220423), the Fundamental Research Funds for the Central Universities (Grant No. DUT21RC(3)032, DUT23YG107 and DUT23YG102), the Application Fundamental Research Project of Liaoning Province (Grant No. 2022JH2/101300259), the Science and Technology Innovation Fund of Dalian (Grant No. 2022JJ12GX011), the China Postdoctoral Science Foundation (Grant Nos. 2022M722999 and 2023T160613), and the National Key R&D Program of China Grant (Grant No. 2022ZD0210700).

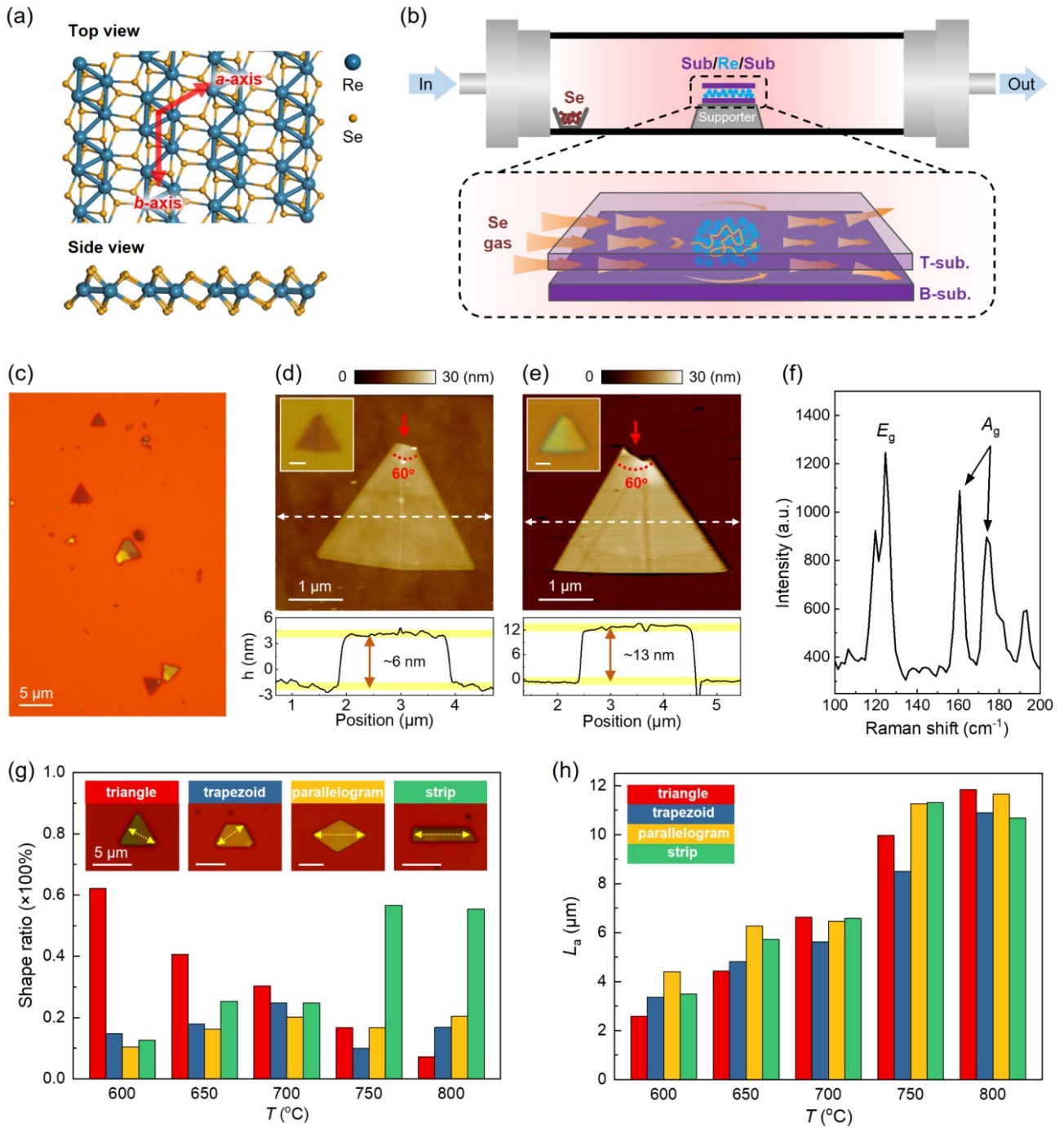

**Figure 1. Space-confined growth of 1T'-ReSe$_2$.** (a) Top and side views of atomic structure of ReSe$_2$ crystal. (b) Schematic of space-confined reactor (SCR) for ReSe$_2$ growth, where the Re and Se powders are used as precursors. The enlarged view shows the gas flow between two substrates. (c) Optical image of SCR ReSe$_2$ flakes grown on SiO$_2$/Si substrate. (d-e) AFM images and height profiles of two triangle-shaped ReSe$_2$ flakes, with corresponding polarized optical images (top insets). (f) Raman spectrum for as-grown ReSe$_2$ sample. (g-h) The effect of growth temperature on the shape and size of ReSe$_2$. (g) Relative quantity proportion for triangle-/trapezoid-/parallelogram-/strip-shaped ReSe$_2$ grown at different temperatures. Top inset: optical images for typical ReSe$_2$ with different shapes, where the yellow double arrows represent the definition of feature sizes ($L$) of these shapes. (h) The temperature dependence of average feature size ($L_a$) for triangle-/trapezoid-/parallelogram-/strip-shaped SCR ReSe$_2$.

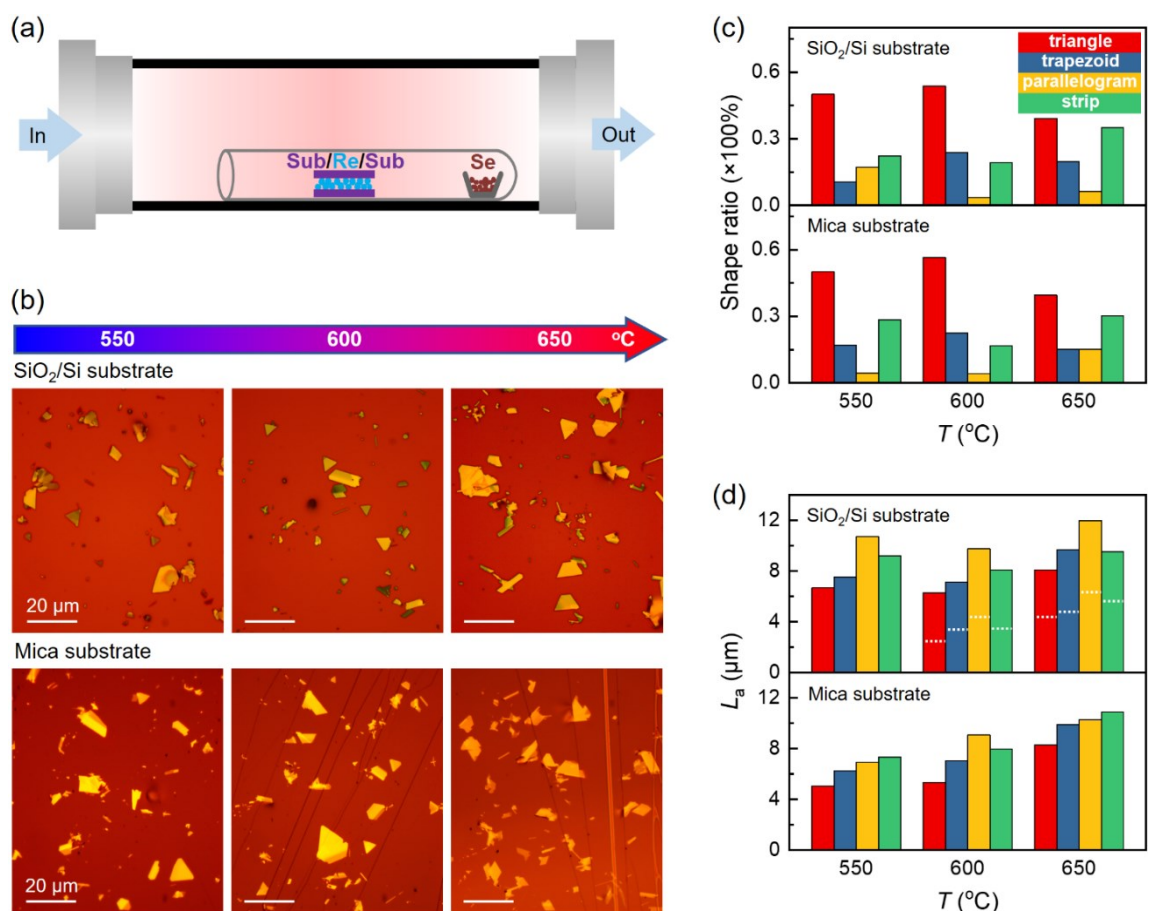

**Figure 2. Low-temperature growth of 1T'-ReSe$_2$ via small-tube-assisted space-confined strategy.** (a) Schematic diagram of small-tube-assisted space-confined reaction (ST-SCR) system. (b) Optical images of ST-SCR ReSe$_2$ grown at 550, 600, and 650 °C on SiO$_2$/Si (top) and mica (bottom) substrates. (c-d) The temperature dependence of (c) relative quantity proportion and (d) average feature size ($L_a$) for triangle-/trapezoid-/parallelogram-/strip-shaped ST-SCR ReSe$_2$. The white dashed lines in (d) represent $L_a$ for SCR ReSe$_2$ in Fig. 1(h).

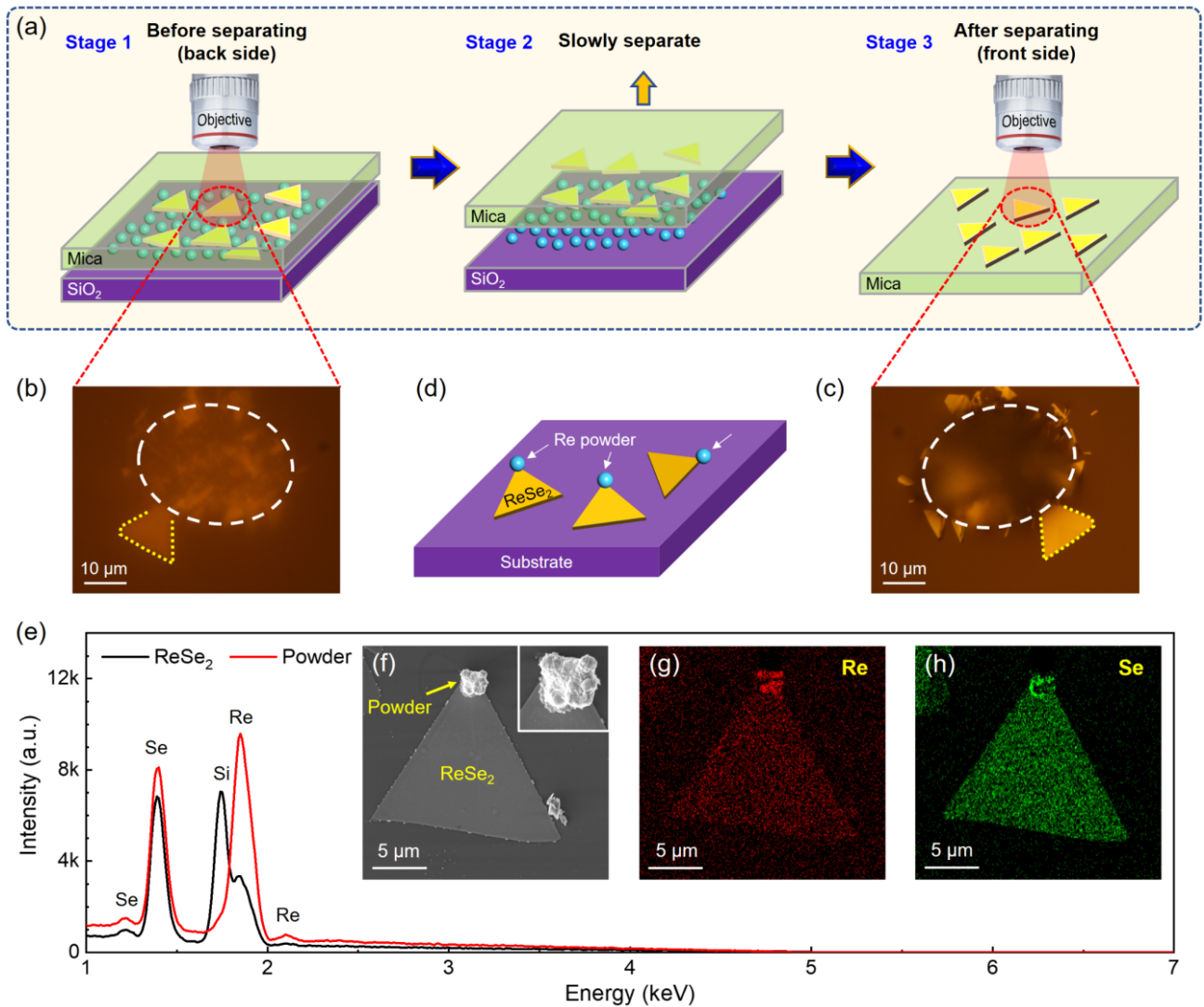

**Figure 3. Relationship between as-grown ReSe$_2$ and solid-phase Re powder.** (a) Schematic of *in-situ* analysis process to observe as-grown ReSe$_2$ samples from back side (before separating) and front side (after separating). (b-c) Optical images of a typical triangle-shaped ReS$_2$ taken from (b) back side and (c) front side. The white dashed circle marks the growth nucleation point covered with residual Re powder. (d) Schematic of relationship between as-grown ReSe$_2$ and solid-phase Re powder. (e) EDX spectra for a triangle-shaped ReSe$_2$ flake grown on SiO$_2$/Si substrate in (f) taken at ReSe$_2$ surface (black curve) and powder position (red curve), with corresponding (g) Re and (h) Se element mapping. Inset in (f): Enlarged SEM image at powder position.

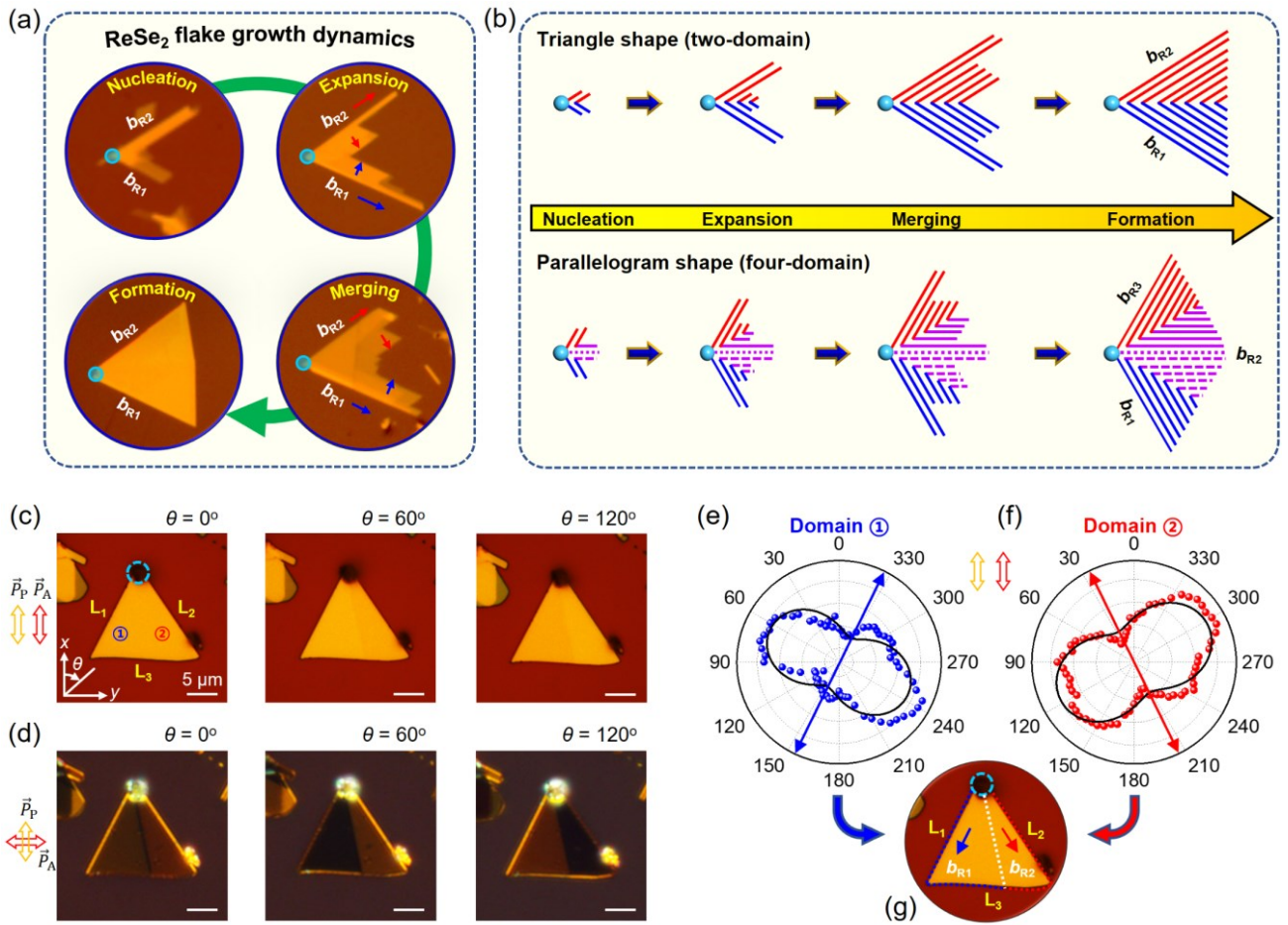

**Figure 4. The solid-phase growth dynamics for two-domain ReSe₂.** (a) The solid-phase-induced four-step growth mechanism for triangle-shaped ReSe$_2$, including nucleation, expansion, merging, and formation, as evidenced using our samples grown on mica substrate. Blue circles point to the Re powder position, red and blue arrows represent the growth directions. (b) Schematics of four-step growth process for triangle-shaped (top) and parallelogram-shaped (bottom) ReSe$_2$, which are composed of two and four domains, respectively. Red, blue, and purple lines represent the *b*-axis directions of different domains. (c-d) Polarized optical images of a triangle-shaped ReSe$_2$ flake with Re powder (blue dashed circle) taken in (c) parallel and (d) vertical configurations with sample rotation angle $\theta$ of 0º, 60º and 120º. Insets: the laboratory coordinate system and the incident light (analyzer) polarization. (e-f) Polar plots of parallel polarized light intensity as a function of sample rotation angle $\theta$ for (e) domain ① and domain ② in triangle-shaped ReS$_2$ in (c), from which (g) the *b*-axis of each domain is determined.

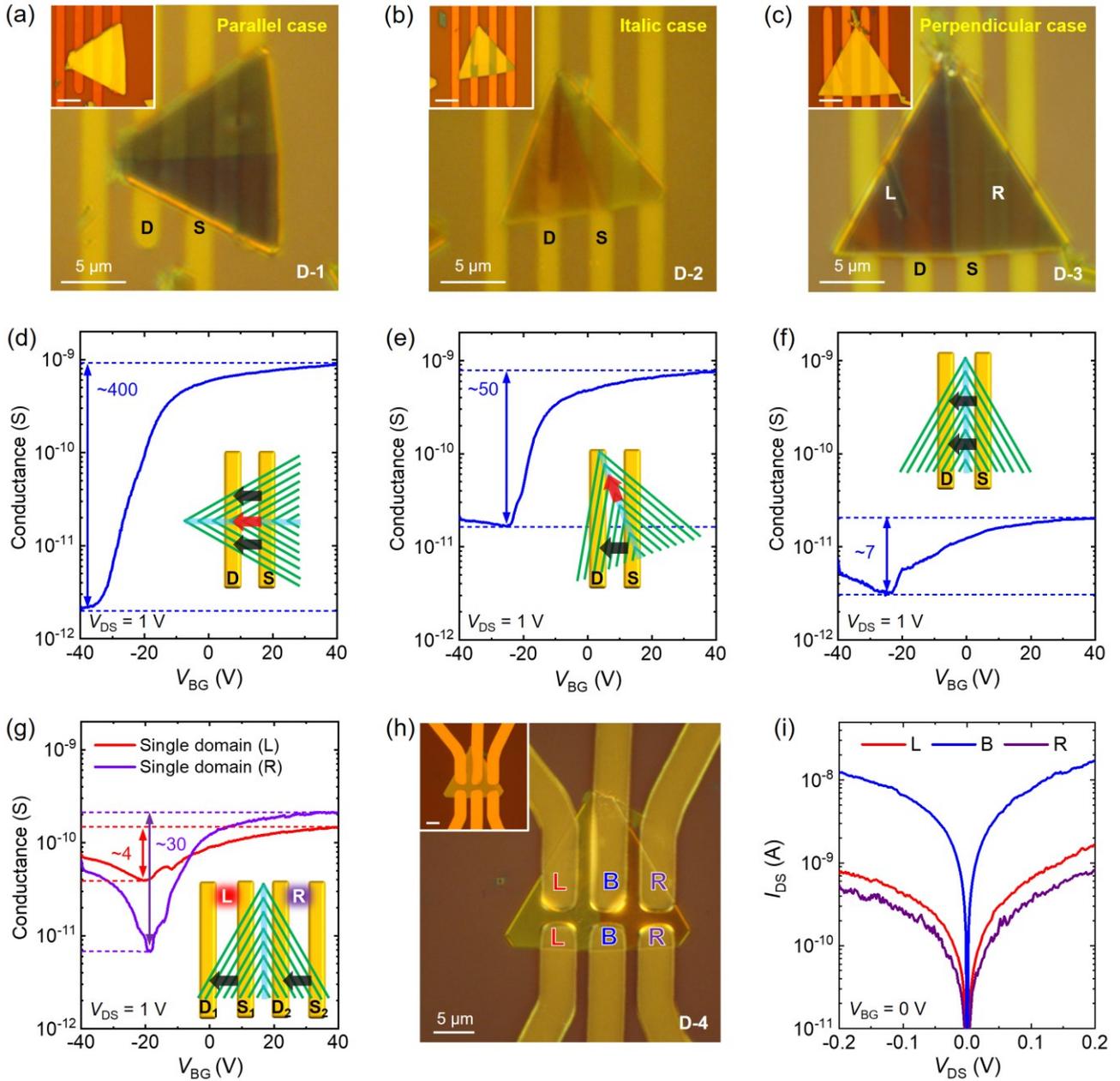

**Figure 5. GB induced modulation of conduction characteristic in two-domain ReSe$_2$.** (a-c) Polarized optical images of three two-domain ReSe$_2$ devices with the GB (a) parallel (D-1), (b) italic (D-2) and (c) perpendicular (D-3) to the channel direction. D and S mark the source and drain electrodes. Insets: the corresponding optical images without analyzer. (d-g) Room-temperature transfer characteristics for sample (d) D-1, (e) D-2, and (f-g) D-3 in (a-c). The double arrows show the conductance switching ratio. Insets: current flow path models, where the black (red) arrows represent the current flow at domain (GB). (h) Polarized and unpolarized (inset) optical images of a two-domain ReSe$_2$ device (D-4) with three pairs of narrow electrodes (L, B, R), which measures the conductance behaviors of two single domains (L, R) and GB (B). (i) Room-temperature $I_{DS}$-$V_{DS}$ ($V_{BG}$ = 0 V) of two single domains (L, R) and GB for sample D-4 in (h) in a log coordinate.
23

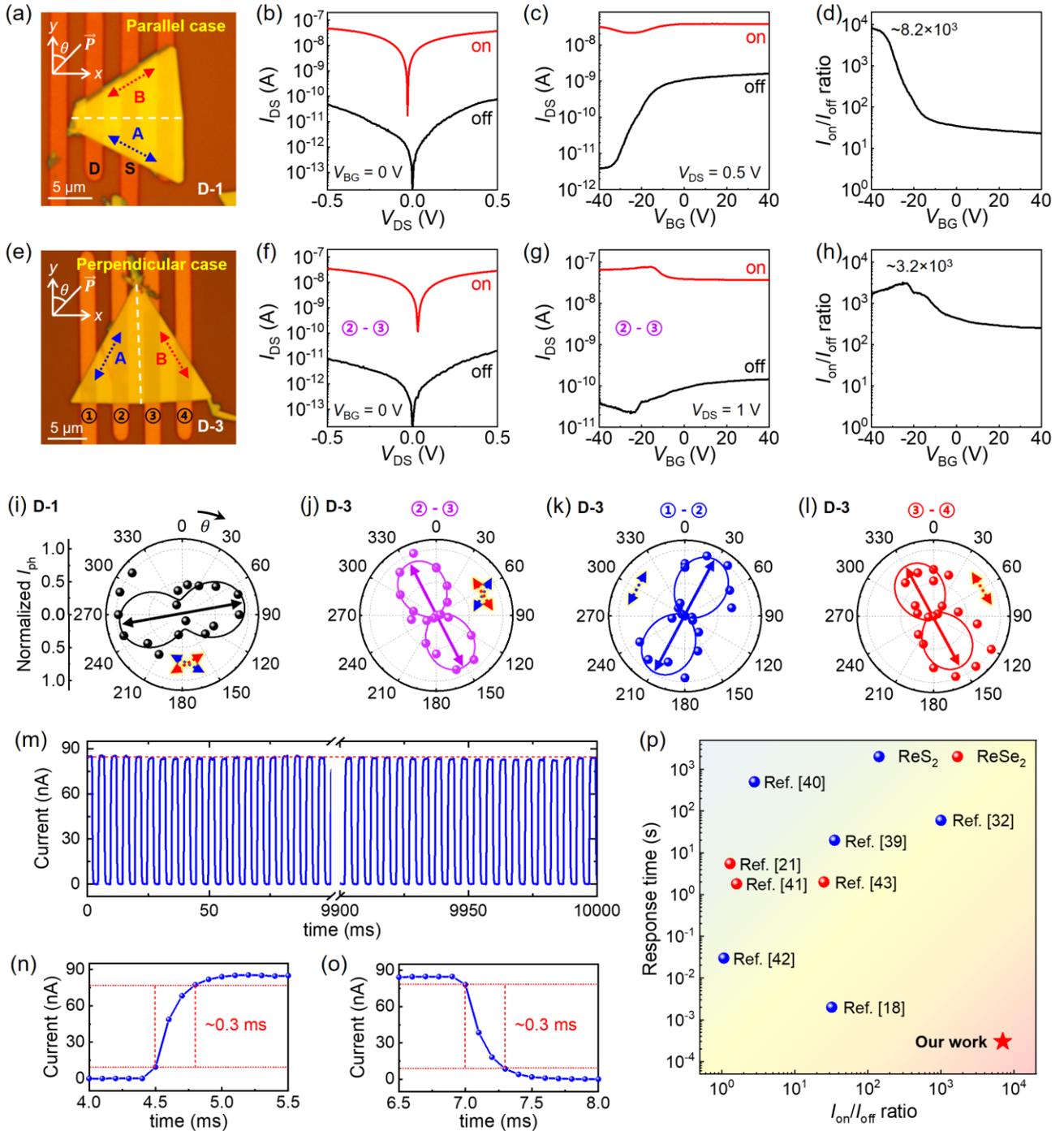

**Figure 6. High-performance of two-domain ReSe$_2$ phototransistors.** (a) Optical image of sample D-1 in Fig. 5(a), with corresponding (b) $I_{DS}$-$V_{DS}$ ($V_{BG}$ = 0 V) and (c) $I_{DS}$-$V_{BG}$ ($V_{DS}$ = 0.5 V) under dark and laser illustration, and (d) $V_{BG}$-dependent current on/off ratio. (e-h) Optical image of sample D-3 in Fig. 5(c), with corresponding (f) $I_{DS}$-$V_{DS}$ ($V_{BG}$ = 0 V) and (g) $I_{DS}$-$V_{BG}$ ($V_{DS}$ = 1 V) under dark and laser illustration, and (h) $V_{BG}$-dependent current on/off ratio. Insets in (a) and (e): the laboratory coordinate system and incident laser polarization. The double arrows show the *b*-axis of each domain, and the white dashed lines mark the GBs. (i-l) Polar plots of normalized photocurrent $I_{ph}$ ($V_{BG}$ = 0 V) as a function of light polarization angle for (i) sample D-1, and channel (j) ②-③, (k) ①-②, (l) ③-④ in sample D-3. (m) Photo-response stability measurement with switching cycles of 2000. The red dashed line indicates ultra-stability of photocurrent amplitude. (n-o) Enlarged view of a (n) rising and (o) descending edge in (m), showing a fast response time of ~0.3 ms. (p) Comparison of response time and $I_{on}/I_{off}$ ratio of our two-domain ReSe$_2$ phototransistors with other reported ReSe$_2$ (ReS$_2$) devices.